# Optical amplification for astronomical imaging at higher resolution

GAL GUMPEL AND EREZ N. RIBAK

*Department of Physics, Technion – Israel Institute of Technology Haifa 32000, Israel*

**Abstract**

Heisenberg's uncertainty principle tells us that it is impossible to determine simultaneously the position of a photon crossing a telescope's aperture and its momentum. Super-resolution imaging techniques rely on modification of the observed sample, or on entangling photons. In astronomy we have no access to the object, but resolution may be improved by optical amplification. Unfortunately, spontaneous emission contributes noise and negates the possible gain from stimulated emissions. We show that it is possible to increase the weight of the stimulated photons by considering photon statistics, and observe an improvement in resolution. Most importantly, we demonstrate a method which can apply for all imaging purposes.

## 1. Introduction

The resolution of an imaging system is limited by diffraction of light crossing the aperture. The diffraction can be explained either classically from the wave nature of light [1, 2] or by taking the quantum approach using Heisenberg's uncertainty principle. However, when quantum correlations are introduced, the classical derivation does not hold anymore, and the quantum approach yields a tighter bound on the diffraction limit [3].

In microscopy, a large variety of techniques changed the field entirely by taking advantage of the quantum nature of light and its interaction with matter. These include, to list but a few, stimulated emission-depletion (STED) and photoactivation localization microscopy (PALM) [4, 5], super-lenses [3] and the recently emerging field of quantum imaging [6-8]. These methods were suggested to overcome the fundamental limit on the resolution of optical microscopy.

Common to all of these methods is the fact that the cost to pay for enhanced resolution is a drop in the number of available photons. Lipson [9] claimed this fact to be a fundamental property of super-resolution imaging. Later Gureyev *et al.* [10] presented a rigorous quantum description of the resolution limit, from which they concluded two important fundamental relations. The first between spatial resolution and noise level ("noise-resolution uncertainty"), which strengthen the claim made by Lipson earlier. The second is a generalized form of the Heisenberg uncertainty principle, in which the uncertainty in measurements of conjugate observables can be traded not only for each other, but also for the signal-to-noise ratio in the measurement of each observable.

The same fundamental limit, arising from Heisenberg's uncertainty relation, still rules optical astronomy. This is mainly because the aforementioned methods either depend on modification of the observed object's emission or require illumination of the object with correlated light, and thus are not applicable in astronomy. An improvement in the resolution of optical astronomy requires a new mindset. Instead of modulating the observed object with a controlled light source, as super-resolution microscopy techniques do, we suggest modulating the light arriving from the observed object, amplified by a controlled active material. The amplification is, of course, raising the noise level as well. Hence, once again the enhanced resolution is traded for poor throughput.

A previous series of papers [11, 12] suggested an improved astronomical resolution, based on heralded photon amplification, performed at the telescope aperture (or its image). The operation of this observation relies on a quantum non-demolition (QND) measurement to trigger a fast detection of the stimulated photon burst, and in that way to rule out most of the spontaneous noise. Numerical simulation of this setup, along with smart post-selection of the data, shows a significant improvement in phase estimation of photons coming from a single star. The improved phase estimation better locates the position of the star with higher precision. However, it is necessary to prove the ability of the system to separate (at least) two unresolved objects. Moreover, the suggested apparatus is not feasible, mainly due to the QND stage that is still a laboratory demonstration, requiring special conditions for operation, such as magnetic or optical trapping, low temperature, and using narrow band stellar light. Also, QND provides partial information about the photon's momentum, since oblique beams take longer to reach the detector.

In this research we suggest and demonstrate a new method to improve the resolution, without the need for a QND stage. Instead, to distinguish between spontaneous and stimulated photons, we narrow the observation band by temporal modulation of the incoming light. In that way, we provide a wide-band all-purpose improved-resolution imaging technique.

## 2. Experiment

A laboratory white-light stable source (Edmund Optics MI-150 fiber optic illuminator) shone on an object, remote from the imaging system. The light was then collected by the objective lens followed by a controlled mechanical chopper which was flipped on and off in periods of ten seconds, thus modulating the incoming signal faster than the medium-term fluctuations of the pump intensity (Fig. 1). The light then passed through a second, collimating, lens to assure a uniform intensity distribution through the amplifier. If the input intensity is not uniform then the amplification will be spatially inhomogeneous, and result in an image with only 'local' resolution enhancements. Another delicate issue regarding the amplification process is the mode competition - the fact that different modes experience amplification in the same gain medium, which leads to cross-saturation effects, where stimulated emission by one mode causes gain saturation for the other modes. Unlike laser operation, here we wish to amplify a wide-band signal so this problem appears also at lower power than in laser cavities. To avoid this issue, the white light source intensity was reduced to a minimum, and placed as far as possible from the objective lens. In addition, the collimated light beam was expanded to the maximum possible width (but still not larger than the active area of the amplifier). With all of these, we could make the photon rate at the camera as low as $10^6$ photons/s/cm$^2$ (which is similar to the astronomical case). Recall that as the lifetime of the excited state according to this spectrum is $\sim 2\cdot 10^{-14}$ s, the amplification process can fairly be considered as a single photon amplification process. In other words, as a result of the amplification, we have stimulated or spontaneous photon clumps. Since the different clumps do not coincide in time or space (because of the low flux), each detected splotch can be considered as a separate event. The addition of tens of thousands of modulated frames, with few events in each, creates a final image.

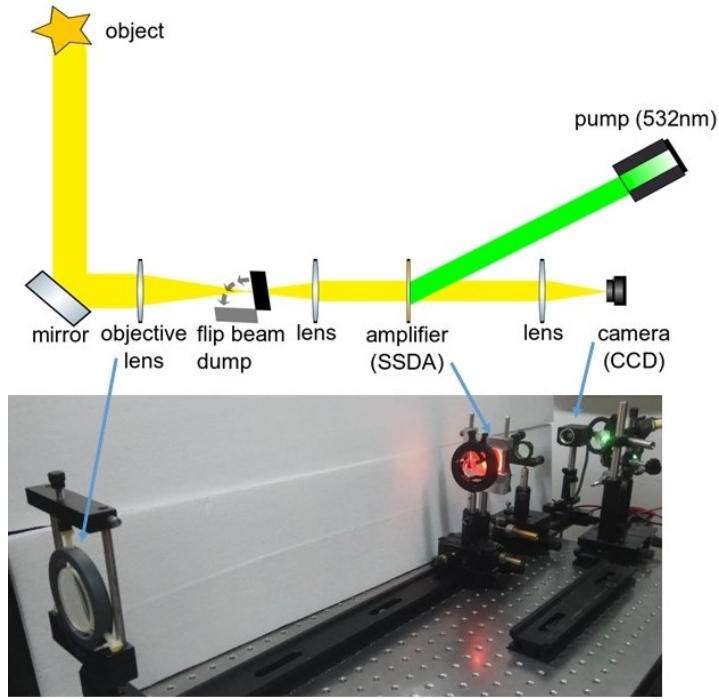

Fig. 1: Experimental setup. The telescope objective ($f/d$ = 200/45 mm/mm) collects the light from the object (2 m away). The second lens ($f/d$ = 50/22) collimates the beam to be of the same size of the active area of the wide-band amplifier. The third lens ($f/d$ = 170/25) focuses the image into a CCD camera (2.2 µm pixels). All lenses are well-corrected achromats. A mechanical chopper (not pictured) can be flipped on and off to modulate the object's light.

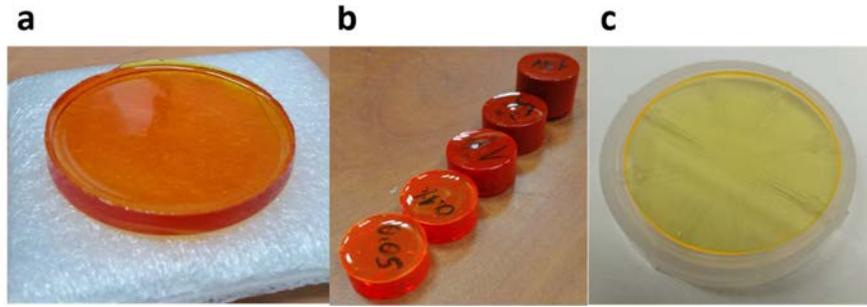

Fig. 2: Solid state die amplifiers. (a) a mold of DCM:PMMA fabricated in the lab, before polishing; (b) Molds of DCM:CR7 fabricated by Shamir-Eyal Ltd.[14]; (c) Glass spin-coated with PMN:PMMA fabricated by F. Vogelbacher at AIT[15]. All samples are about 5 cm in diameter

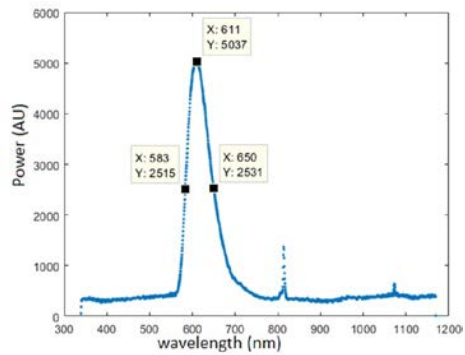

Fig. 3: Fluorescence spectrum of DCM:PMMA.

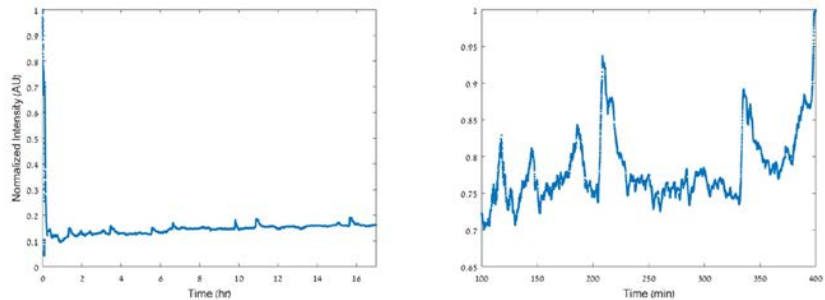

Fig. 4: Pump intensity during operation. Left: the entire time. Right: Zoom in on 5 hours starting 100 minutes after the pump was turned on. The modulation was faster than the spikes frequency.

The last lens finally collects the light to form an image of the source on the CCD camera. Both the camera and the mechanical chopper were connected to a PC and controlled by Matlab software. The acquisition algorithm consists of capturing a 10 ms image, blocking the beam and capturing another 10 ms image. All signal and background images were averaged and subtracted to yield a single difference image.

As an amplifying agent, we chose a solid-state dye (SSDA, Fig. 2). This is essentially a dye-doped polymer. It was first suggested as a gain medium for laser applications by MacFarland and Soffer [13]. It was chosen for its ease of handling and spectrum. Pumping the SSDA at the right frequency will excite the dye molecules and cause them to fluoresce. If another light beam, at the decay frequency of the dye, will cross the active (excited) section it will cause a stimulated emission. This is the act of amplification.

The first samples consisted of 4-(Dicyanomethylene)-2-methyl-6-(4-dimethylaminostyryl)-4H-pyran (DCM) dye doped in a Poly(methyl methacrylate) (PMMA) matrix (abbreviated DCM:PMMA). A DCM powder was first mixed with MMA using a magnetic stirrer until it appeared as a uniform reddish liquid. Then a photoinitiator (Irgacure 124) was added to the mixture and stirred again. When the liquid mixture became uniform it was molded into a Petri dish and put under UV-light for few days to solidify. The sample was then polished to optical quality using a CNC machine. Other samples were fabricated with Polychloroprene (or Chloro-prene Rubber — CR7) instead of PMMA. These samples were solidified in an oven. The second process took place in Shamir-Eyal Ltd.'s fabrication labs [14]. Later versions were fabricated using 2-(4-(bis(4-(tert-butyl)phenyl)amino) benzylidene)malononitrile (PMN) dye whose response was found to be less sensitive to spatial changes in the concentration of the dye in the polymer [15]. Unlike the molding process, the PMN:PMMA was spin-coated over a glass disk. The final optical quality of the SSDA was limited, and we were somewhat below the theoretical diffraction limit of the set up. Without measuring the wave front error, we can roughly estimate it from the final CTF to be up to 100 nm.

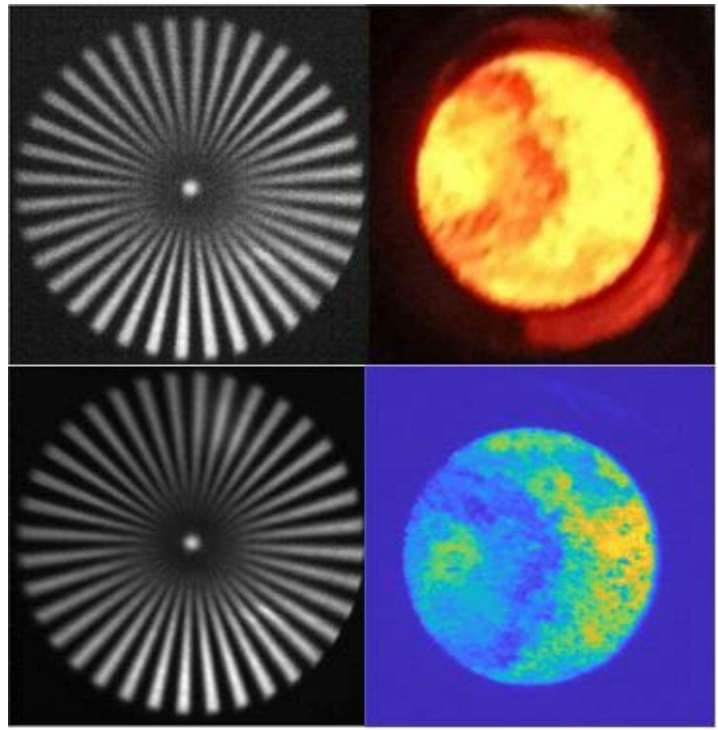

Fig. 5: Left panels: An image of a 36-spoke wheel of diameter 1 cm, and below its amplified image on the telescope's camera. Notice the slight bending of the spokes near the center, resulting from weak aberrations. Right: A direct image of a fiber bundle output of diameter 5 mm, consisting of single-mode fibers with diameter of about 50 microns, above its amplified image.

Absorption and emission spectra of DCM are generally in the ranges of 450-550nm and 600-700nm, respectively [16]. The exact spectrum is slightly modified depending mainly on the solvent. The fluorescence spectrum of the DCM:PMMA sample, under 10mW green laser (532nm) pump, was measured in a spectrometer (Fig. 3). We observed a peak at 611nm with FWHM of 70nm, which is significantly wider than other quantum amplifiers (but still not covering the whole visible spectrum).

The pump laser was a common frequency-doubled Nd:YVO4 diode-pumped solid state laser pointer. The laser pointer was connected to a stabilized DC source of up to 5 V and generated an output power of up to 33 mW, where 98% of the power was absorbed by the SSDA and the rest reflected off its surface.

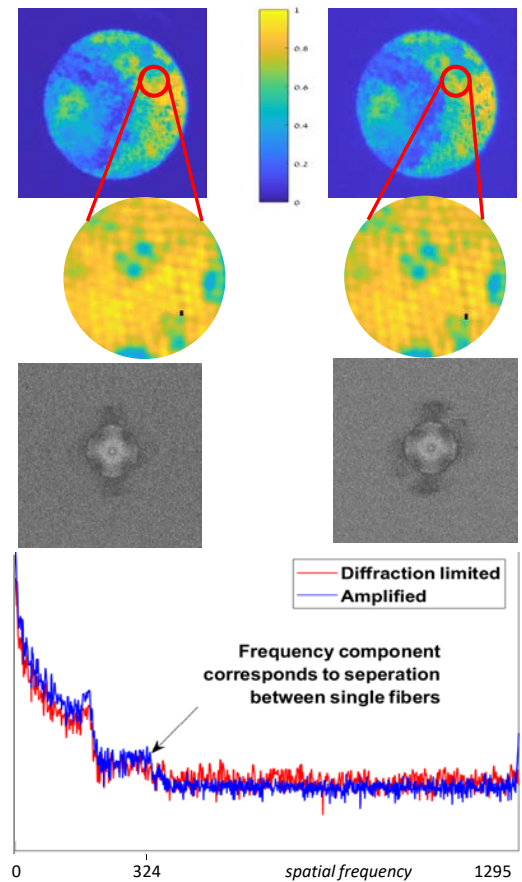

Fig. 6: Resolution comparison for real-world objects. Left: regular imaging. Right: Amplification imaging, both with the same number of frames. The resolution-size bar on the close-up images is placed at the same position to emphasize the enhanced separation between single fibers. Middle row: Fourier transforms of the object images in the top row; Bottom row: vertical cuts through the transform for the two cases.

The pump intensity was not stable due to changes in temperature during its operation. A measurement of the time-dependent intensity (Fig. 4) revealed some expected and unexpected fluctuations. First, we see a fast decrease in the initial intensity taking about 15 minutes. Another noticeable phenomenon is the long-term drift in the average intensity. In addition, at the right panel of Fig. 4, it appeared that in periods of ten minutes there were fluctuations up to 15% of the average intensity. These medium-term fluctuations are not negligible, and we had to modulate the signal faster than this period in order to filter them out.

## 3. Results

To visualize the improvement in resolution achieved by the new method we acquired images of a real-world object. The object used was the output of a fiber bundle illuminated by a thermal light source (Fig. 5). In Fig. 6 we show the reconstructed images of the object with and without amplification. Both images were averaged over 20,000 frames. The close-up shows an area of dense fibers, where the improvement in separation capability clearly appears.

The images in the middle row are the Fourier spectra of the images above them. As the system resolution improves, tinier details should appear in the image, giving rise to higher frequencies in the Fourier domain. It is quite challenging to see this effect in the 2D Fourier image, but looking at the 1D cross-section at the bottom confirms that there is a rise in the Fourier amplitude at the frequency that corresponds to the separation between single-mode fibers. The height of the image transform is 2592 Fourier frequencies, whereas the separation between single-mode fibers is about 8 pixels, which corresponds approximately to the 324th Fourier component. While the improvement is clear, we wished to employ a more standard method.

A common evaluator of the resolution of an optical system is the modulation transfer function (MTF). The MTF of an optical system is a measurement of its ability to transfer contrast at a particular spatial frequency from the object to the image. The spatial frequency is inversely proportional to the separation between points on the target, therefore when we come closer to the resolution limit of the optical system, the MTF decreases. At exactly the resolution limit

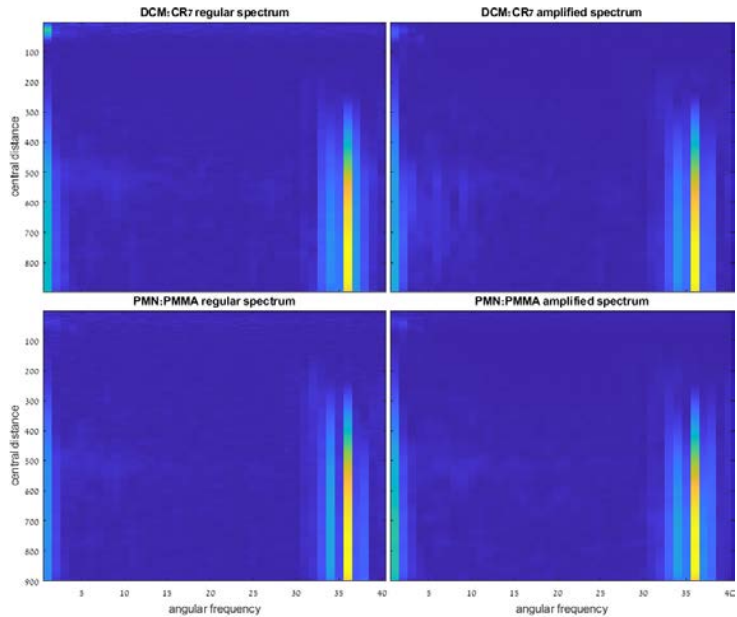

Fig. 7: The angular modulation spectrum at different radii for regular and amplification imaging using both DCM:CR7 (top) and PMN:PMMA (bottom) amplifiers, obtained after converting the spoke wheel pattern from polar to cartesian coordinates. With amplification, the dominance of the 36[th] cycle is clearer. The radii are given (in pixels) from the wheel center down to the spokes' edges. All panels share the same frequency scale.

(also referred to as 'cut-off frequency') and beyond it, the MTF drops to zero.

The MTF can be calculated directly by measuring the contrast of sine-wave targets at different spatial frequencies. However, for a square-wave target the contrast measurements result in the equivalent contrast transfer function (CTF), which has been previously shown to be related to the MTF through Coltman's formula [17]. In these measurements we used a spoke wheel target consisting of 36 equally spaced spokes (Fig. 5). The advantage of this type of target is that it allows to capture many spatial frequencies within a single image, by measuring the modulations around concentric circles at different radii. The spatial frequency is inversely proportional to the radius. The modulated intensity curve at each radius was decomposed into its angular Fourier components, then the component corresponding to 36 cycles was reconstructed, and its contrast measured. A set of these contrast measurements forms the CTF, which is a sufficient approximation of the MTF when measured at frequencies greater than one-third of the cut-off frequency [18, 19].

In Fig. 7 we show the normalized Fourier spectra of the modulations *around* circles with different radii. The two rows are for the solid-state dye amplifiers we used, DCM:CR7 and PMN:PMMA. We can clearly see that using the amplification imaging method, the peak at a value of 36 cycles per circle is more prominent. This fact suggests we have achieved an improvement in the resolution. However, our optical system had small aberrations, which means that the spokes were not straight all the way to the center (Fig. 5). That means that their angular frequency slightly deviated from 36, and to reduce that effect we included the spectrum of frequencies 34 to 38. This addition is still sensitive to the presence of the spokes at nearby frequencies, until they sink below the noise level. To measure this noise level, we took the mean level nearest to the center, where clearly there was no signal (central distance 60-100 in Fig. 7).

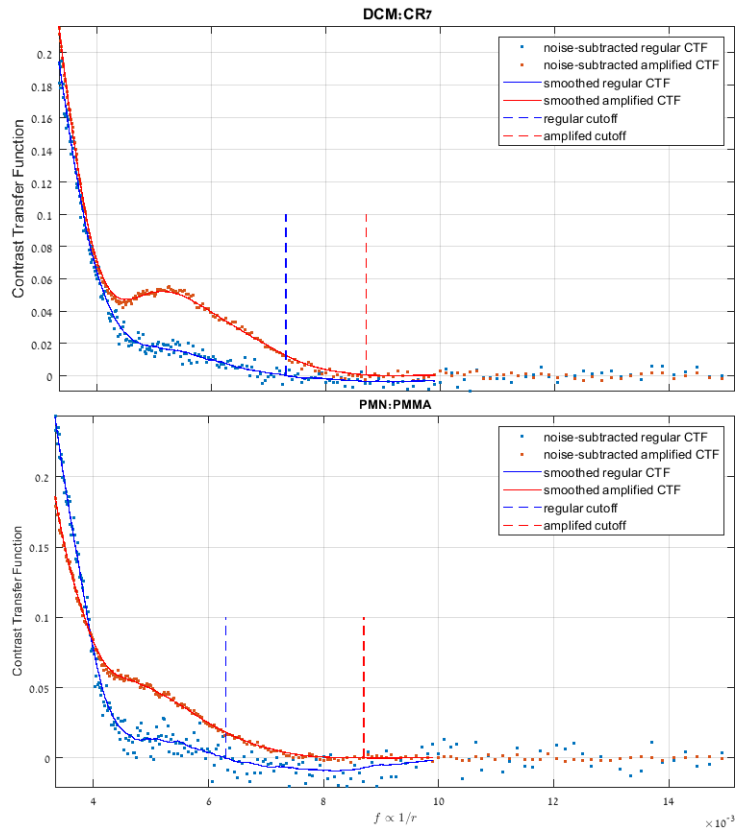

Fig. 8: Reconstructed contrast transfer curves for regular and amplification imaging using DCM:CR7 and PMN:PMMA amplifiers. The single dots are smoothed into curves showing the CTF values. The zero CTF is set according to the mean noise beyond the cut-off, as marked by the vertical broken lines.

In Fig. 8 we evaluated the MTF curves through the CTF at higher frequencies (that is, nearer to the spoke center). The CTF curves achieved by using the two different gain media are compared to the CTF curves achieved by regular imaging of the same objects. The CTF of the DCM and PMN dyes were averaged over four data-sets consisting of 20,000 frames each. The CTF errors (not marked) were of order 0.005, from the standard deviation of the data. The cut-offs are the frequency at which the data points drop to the background level. That level was the average of the spectrum nearer to the center of the spoke wheel, well beyond the cut-off frequency (here above $10^{-2}$ pixel$^{-1}$). For visualization, we subtracted it to see where the CTF drops to zero. For the DCM-dye amplifier, the cut-off was measured to be $(87 \pm 2) \times 10^{-4}$ pixels$^{-1}$ compared to $(73 \pm 2) \times 10^{-4}$ pixels$^{-1}$ for regular imaging. This corresponds to an improvement of $(19 \pm 1)\%$ in the CTF. For the PMN-dye amplifier, the improvement is even higher as the cut-offs were at $(87 \pm 2) \times 10^{-4}$ pixels$^{-1}$ using the amplification method, compared to $(63 \pm 2) \times 10^{-4}$ pixels$^{-1}$ with regular imaging. This corresponds to an improvement of $(38 \pm 2)\%$ in the CTF.

## 4. Conclusions

The results of this experiment provide an evidence for improvement of the spatial resolution with optical amplification of thermal light. The two leading approaches in modern super-resolution imaging are sub-diffraction fluorescence microscopy (e.g. STED and PALM) and quantum entanglement (e.g. from spontaneous parametric down-conversion). The former is well-established and achieves an order-of-magnitude improvement in the resolution, which goes well beyond the achievement we presented. On the other hand, these techniques are more complicated and limited only to bio-imaging applications. The latter, quantum entanglement, was demonstrated only recently where the state-of-the-art setup achieved 17% improvement in the MTF [20]. This approach can be applied to a broader range of imaging applications, but is still limited to the field of microscopy because it has to control the illumination of the object. Traditional passive methods, such as Toraldo phase rings [21], were shown experimentally to increase the resolution by 19%, but are limited to isolated objects [22]. Classical post-detection super-resolution, such as by Airy disk deconvolution or by using object priors, improves resolution for images with good photon statistics. But, using the same initial photon statistics, we were able to reach finer details only when we added amplification.

The novelty of the amplification imaging method lies in its simplicity, its ability to reach modern state-of-the-art achievements, and mainly in the fact that it is a method which can be applied for all passive imaging purposes, including astronomy. We were able to increase the resolution, despite the fact that our set-up did not use photon-tagging equipment, and was slightly below the theoretical diffraction limit. Application of adaptive optics, routinely used for reducing temporal and permanent aberrations in astronomy, can improve the resolution even beyond the theoretical diffraction limit.

For future progress it would be interesting and important to study the limitations on this method, both theoretically and experimentally, with an emphasis on the trade-off between improvement in resolution and loss of photons. Specifically, we would like to explore new ways to achieve higher gain levels and wider amplification bands.

## 5. Acknowledgments and disclosures

*Acknowledgements*
The authors would like to thank Shamir-Eyal Ltd. for fabrication of DCM:CR7 samples, and F. Vogelbacher and colleagues for fabrication of PMN:PMMA samples. Also, thanks to S. G. Lipson and B. M. Levine at the Technion, and the astronomy group at the University of Sydney for useful discussions.